\documentclass[twocolumn,showpacs,preprintnumbers,amsmath,amssymb,prl,superscriptaddress]{revtex4}

\usepackage{graphicx}

\newcommand{\expct}[2]{\left\langle #1 \right\rangle_{#2}}
\newcommand{\expcts}[2]{\langle #1 \rangle_{#2}}
\newcommand{\cumu}[2]{\langle\!\langle #1 \rangle\!\rangle_{#2}}
\newcommand{\cumus}[2]{\langle\!\langle #1 \rangle\!\rangle_{#2}}
\newcommand{\C}{\mathcal{C}}
\newcommand{\TC}{T_{\C}\, }
\newcommand{\T}{\mathcal{T}}

\begin{document}

\title{Charge transfer statistics of a molecular quantum dot\\with a
vibrational degree of freedom}
\author{T.~L.~Schmidt}
\affiliation{Departement Physik, Universit\"at Basel,\\
 Klingelbergstrasse 82, CH-4056 Basel, Switzerland}
\author{A.~Komnik}
\affiliation{Institut f\"ur Theoretische Physik, Universit\"at Heidelberg,\\
 Philosophenweg 19, D-69120 Heidelberg, Germany}
\date{\today}

\begin{abstract}
We analyze the full counting statistics (FCS) of a single-site
quantum dot coupled to a local Holstein phonon for arbitrary
transmission and weak electron-phonon coupling. We identify
explicitly the contributions due to quasielastic and inelastic
transport processes in the cumulant generating function and
discuss their influence on the transport properties of the dot. We
find that in the low-energy sector, i.e.~for bias voltage
 and phonon frequency much smaller than the dot-electrode contact
 transparency, the inelastic term causes a sign change
in the shot noise correction at certain universal values of the
transmission. Furthermore, we show that when the correction to the
current due to inelastic processes vanishes, all the odd order
cumulants vanish as well.
\end{abstract}


\maketitle


During the past decade molecular electronics has evolved into an
important branch of condensed matter physics \cite{cuniberti}.
Nowadays it is possible to electrically contact molecules of
almost any geometry and complexity ranging from carbon nanotubes
to hydrogen molecules
\cite{Tans,LeRoy,Smit,kiguchi08,djukic05,djukic06}.
Moreover, the experimental observables are not restricted to the
linear conductance properties any more, but encompass also the
nonlinear current-voltage characteristic $I(V)$ as well as the
shot noise \cite{kiguchi08}. Especially in the case when
vibrational degrees of freedom are involved, these transport
quantities display a number of very interesting features
\cite{PhysRevLett.88.226801,PhysRevLett.92.206102,
PhysRevLett.93.266802,sapmaz:026801,ParkMcEuen,ParkRalph,YouNatelson,Pasupathy}.

One of the most pronounced effects of the interaction between the
electronic and vibronic degrees of freedom is the abrupt change of
the system conductance once the applied voltage is increased
beyond a threshold which is related to the excitation energy of
molecular vibrations
\cite{vega:075428,paulsson:201101,PhysRevB.68.205406,egger:113405,
Smit,PhysRevLett.88.226801,ParkMcEuen,Pasupathy,LeRoy}. This is
the reason why measurements around this turning point have become
an invaluable instrument for the experimental investigation of
such systems. Interestingly, the sign of this conductance step
depends crucially on the junction transparency, being negative for
almost perfect transmission and positive in the opposite case.
Most theoretical efforts have been centered around these two
limiting cases
\cite{paulsson:201101,PhysRevLett.92.206102,vega:075428,galperin06,egger:113405,koch05,koch05_2,flindt05}.
According to Ref.~\cite{egger:113405}, the transition is
nonuniversal and the precise condition involves all system
parameters \cite{glazman88}. In general it occurs at an
intermediate value of the transmission.

Thus far mainly the nonlinear current-voltage characteristic has
been analyzed in this regime. However, for future applications it
is of importance to also possess information about the noise
properties of such systems \cite{galperin06,kiguchi08}. A very
convenient tool to calculate a variety of transport properties is
the full counting statistics (FCS) which gives the probability
distribution $P(Q)$ to transfer $Q$ elementary charges during a
fixed (very long) waiting time $\T$. The average value
$\cumus{Q}{}$ is then directly related to the current and its
variance $\cumus{Q^2}{}$ to the noise power. The relevance of
higher order cumulants $\cumus{Q^n}{}$ has been demonstrated in
Ref.~\cite{reulet}. Moreover, the analytic structure of the
cumulant generating function $\ln \chi (\lambda)$ can give
invaluable insights into the nature of the processes contributing
to the transport
\cite{reznikov,muz,weisssaleur,cuevas,AndersonFCS}. That is the
reason why we would like to address the FCS of the molecular
quantum dot coupled to a Holstein phonon.

We model the system by the following Hamiltonian,
\begin{equation}
 H = H_L + H_R + H_d + H_T + H_{ph} + H_{el,ph} \, .
\end{equation}
The terms $H_{L,R}$ describe the left and right electrodes
in the language of the respective electron field operators
$\psi_{L,R}(x)$. We model them as noninteracting fermionic
continua held at the chemical potentials $\mu_{L,R}$. The applied
bias voltage is then given by $V=\mu_L - \mu_R>0$ (we use units
where $e=\hbar=1$). The particle transport between the dot and the
electrodes is mediated by a local (symmetric \footnote{The contact
asymmetry can easily be taken care of. We did not include it
explicitly since it does not generate qualitatively new features but
leads to cumbersome calculations.}) tunneling coupling at $x=0$
with an amplitude $\gamma$
\begin{equation}
 H_T = \gamma \, d^\dag \left[ \psi_L(x=0) +
    \psi_R(x=0)  \right] + \mbox{h.c.} \, ,
\end{equation}
where $d$ is the annihilation operator of the electron on the dot.
We assume it to be spinless and modeled by $H_d = \Delta d^\dag
d$, $\Delta$ being the bare energy of the dot state which can be
tuned by an applied gate voltage. The local (single-mode) phonon
is described by $H_{ph} = \Omega a^\dag a$ while the
electron-phonon coupling term with the amplitude $g$ is given by
\begin{equation}
 H_{el,ph} = g\ q\ d^\dag d \, ,
\end{equation}
where $q = a + a^\dag$ is proportional to the phonon displacement operator.

In order to determine the FCS, we calculate the cumulant
generating function (CGF) $\ln \chi(\lambda) = \ln \expcts{e^{i
\lambda Q}}{}$ which gives access to the cumulants (or irreducible
moments) $\cumus{Q^n}{}$ of $P(Q)$. The formalism for CGF
calculation has been developed in Refs.~\cite{lll,nazarovlong} and is by
now well adapted to quantum impurity problems
\cite{bagrets06,AndersonFCS,HBTLetter}. We chose to use
perturbation theory in the electron-phonon coupling $g$. The
corresponding CGF then reads
\begin{equation}
 \ln \chi(\lambda) = \ln \chi_0(\lambda) + \ln \chi'(\lambda) \, .
\end{equation}
The first term is the CGF of the clean system at $g=0$. At zero temperature,
it is known to be given by
\begin{align}
 \ln \chi_0(\lambda) =\T \int_{-V/2}^{V/2}
 \frac{d\omega}{2\pi} \ln \left[ 1 + T_0(\omega) (e^{i\lambda} - 1)
 \right]\, ,
\end{align}
where $T_0(\omega) = \Gamma^2 /[(\omega - \Delta)^2 + \Gamma^2]$
is the single-particle transmission coefficient of the resonant
level model \cite{AndersonFCS,dejong} and $\Gamma = 2 \pi \rho_0
\gamma^2$ is the dot-electrode contact transparency, which in the
wide flat band model depends on the energy independent density of
electronic states in the electrodes $\rho_0$. The correction is
treated in the spirit of Ref.~\cite{schmidt07_4} which is based on
the generalized Keldysh approach proposed in Ref.~\cite{reznikov}.
It is given by
\begin{equation}
 \chi'(\lambda) = \expct{ \TC \, e^{-i g \int_\C ds\ q(s)\ d^\dag(s)
d(s)} }{\lambda},
\end{equation}
where the expectation value is taken with respect to the
noninteracting Hamiltonian $H_L + H_R + H_d + H^\lambda_T +
H_{ph}$ and is time-ordered on the Keldysh contour $\C$. The
dependence on the counting field $\lambda$ is contained in
$H_T^\lambda = \gamma d^\dag \left( e^{i \lambda} \psi_L + \psi_R
\right) + \mbox{h.c.}$ The counting field has different signs on
the two Keldysh branches, $\lambda(t) = \pm \lambda$ for $t \in
\C_\mp$.

An application of the standard linked cluster expansion generates
two different contributions to the lowest-order ($g^2$) term: $\ln
\chi' = \ln \chi_1 + \ln \chi_2$. The first one is present only if
$\Delta \neq 0$, and is thus a consequence of a detuning of the
dot from the particle-hole symmetric point,
\begin{eqnarray}\label{chi1}
 \ln \chi_1(\lambda) & = & -\frac{i g^2 \T}{2} \sum_{k,l=\pm} (kl)
\int
 ds_1\ ds_2\ A^{kl}(s_1-s_2)
 \nonumber \\
& \times &
 N^k(s_1) N^l(s_2).
\end{eqnarray}
Here, $A(t,t') = -i \expcts{\TC q(t) q(t')}{0}$ denotes the
phonon Green's function (GF) in Keldysh space and $N^k(s) =
\expcts{d^\dag(t_k) d(t_k)}{\lambda} - 1/2$ for $t_k \in \C_k$,
$k=\pm$ are generalized dot population probabilities. Due to the
explicit $\lambda$-dependence on the Keldysh branch $k$ they are
different on the forward/backward path $\C_k$.

The other contribution is given by
\begin{equation}\label{deflnchi2}
\ln \chi_2(\lambda) = -\frac{g^2 \T}{2}
\sum_{k,l=\pm}  (kl) \int \frac{d\omega}{2\pi}
 \pi^{kl}(\omega) A^{lk}(\omega) \, ,
\end{equation}
where the generalized charge polarization
loops are defined in terms of the dot level GFs
$D(t)=-i \expcts{\TC d(t) d^\dag(t')}{\lambda}$ by
\begin{equation}\label{defpi}
 \pi^{kl}(\omega) = i \int\frac{d\omega'}{2\pi} D^{kl}(\omega+\omega') D^{lk}(\omega')
 \, .
\end{equation}
Both contributions (\ref{chi1}) and (\ref{deflnchi2}) can be
calculated using the unperturbed dot GFs \cite{AndersonFCS}:
\begin{widetext}
\begin{align}               \label{Ds}
    D(\omega)
&=
    \left(
      \begin{array}{cc}
        D^{--} & D^{-+} \\%
   D^{+-} & D^{++}
      \end{array}\right)
= \frac{1}{\mathcal{D}_0(\omega)}
    \left(
      \begin{array}{cc}
        (\omega-\Delta) + i \Gamma (n_L + n_R - 1) &
        i \Gamma ( e^{ i\lambda/2} n_L + e^{-i\lambda/2} n_R) \\
        i \Gamma [ e^{-i\lambda/2} (n_L - 1) + e^{ i\lambda/2} (n_R -
1)]&
        -(\omega-\Delta) + i \Gamma (n_L + n_R - 1) \\
      \end{array}
    \right) \, ,
\end{align}
\end{widetext}
where the denominator is given by
\begin{align}
     \mathcal{D}_0(\omega)
&=
    (\omega-\Delta)^2 + \Gamma^2 \Big[
    n_L(1-n_R) ( e^{i\lambda} - 1) \notag \\
&+ n_R(1-n_L) (e^{-i\lambda} - 1 ) +
1\Big] \, .
\end{align}
In the wide flat band model the electrodes are noninteracting
fermions with Fermi distribution functions $n_{R,L}(\omega)$. The
chemical potentials in the left/right electrode are $\pm V/2$.
From now on we concentrate on the zero temperature results. The
presented results hold for all temperatures much smaller than the
smallest energy scale in the system. The phonon GFs are given by
$A^{--(++)} (t) = - i \exp( \mp i\Omega|t|)$ and $A^{-+(+-)} (t) =
- i \exp( \pm i\Omega t)$. Equation (\ref{Ds}) allows one to
calculate the generalized population probabilities $N^k$ and one
finds from Eq.~(\ref{chi1})
\begin{align}                 \label{xi1}
\ln \chi_1(\lambda)
&=
    -\frac{g^2 \T}{2 \pi^2 \Omega} \,
    \ln \left[ \frac{f(V/2)}{f(-V/2)} \right]
 \notag \\
&\times \sum_{p=\pm}
    p \, \arctan\left(\frac{V/2 - p \, \Delta}{\Gamma}\right) \, ,
\end{align}
where we defined a $\lambda$-dependent function $f(\omega) =  [1 +
T_0(\omega) (e^{i \lambda} -1)]^{-1}$. The contribution $\ln
\chi_2$ can be written as a sum of two terms which describe
\emph{quasielastic} and \emph{inelastic} processes
\cite{egger:113405}. The quasielastic part reads
\begin{equation}                    \label{xiqel}
 \ln \chi_{qel}(\lambda) = - \frac{\T}{2 \pi} \int_{-V/2 -
 \Delta}^{V/2-\Delta} d \omega \frac{\omega}{\omega^2 + \Gamma^2
 \, e^{i \lambda}} \Sigma^R_R(\omega) \, ,
\end{equation}
where
\begin{align}
 \Sigma^R_R(\omega)
&=
 \sum_{k,l=\pm} \frac{g^2 \, \Gamma}{(\omega + k \Omega)^2 + \Gamma^2}
  \\
&\times
 \left\{ \frac{\omega + k \Omega}{2\Gamma} \left[ 1 + \frac{2 k}{\pi}
 \arctan\left(\frac{l V/2 - \Delta}{\Gamma}\right) \right]
 \right. \notag \\
&\left.
 + \frac{k}{\pi} \ln\left[\frac{\sqrt{(l V/2 - \Delta)^2
 + \Gamma^2}}{|\omega + k \Omega -(l V/2 -\Delta)|} \right] \right\}
 \notag
\end{align}
is the real part of the retarded dot self-energy
$\Sigma^R(\omega)$ \cite{egger:113405}. It is related to the
formation of the phonon sidepeak in the spectral function of the
dot and can be considered as a renormalization of the bare
transmission.

The other contribution is the \emph{inelastic} one\footnote{In the
related study \cite{ALY} the inelastic components are defined
differently.}. It only contributes for $V
> \Omega$ and is given by
\begin{align}                 \label{xiinel}
\ln \chi_{inel}(\lambda)
&=
 \frac{g^2 \T}{2 \pi} \theta(V-\Omega) \int_{-V/2-\Delta}^{V/2-\Delta}
 d\omega
 \\
&\times
 \sum_{k=\pm} \theta(V/2 -k \Delta - k \omega - \Omega)
 \frac{1}{\omega^2 + \Gamma^2 e^{i\lambda}}
 \notag \\
&\times
  \left[
 \frac{\omega (\omega + k\Omega)}{(\omega + k\Omega)^2 + \Gamma^2}
 -
 \frac{1}{2} \frac{\omega (\omega + k\Omega)
 - \Gamma^2 e^{i\lambda}}{(\omega + k \Omega)^2 + \Gamma^2 e^{i\lambda}}
 \right]\, . \notag
\end{align}
Equations (\ref{xi1}), (\ref{xiqel}) and (\ref{xiinel}) are the
main result of this paper. Compatible results were obtained
independently in two related studies \cite{ALY,BelzigPhononFCS}.
They represent the full perturbative result for the correction to
the CGF due to the phonon at arbitrary transmission and are valid
at zero temperature and for small $g$ \footnote{The complete CGF
should satisfy $\ln \chi(\lambda=0) = 0$, so one has to subtract
the $\lambda=0$ term from the above results. However, we are
interested mainly in the cumulants which are proportional to the
derivatives with respect to $\lambda$. That is why we ignore this
constant.}.

The term $\ln \chi_1$ is due to
the renormalization of the level energy $\Delta$ by the presence of
the phonon
\cite{egger:113405}, and is the leading term for small $\Omega$.
This is precisely the condition when the adiabatic (or
Born--Oppenheimer) approximation is valid.
The interpretation of other contributions is more lucid in the low
energy sector when $\Omega, V \ll \Gamma, \Delta$. Then we obtain
\begin{eqnarray}                  \label{nr1}
  \ln \chi_1(\lambda) &=& \frac{2 g^2 \T \Delta V}{\pi^2\Omega}
\arctan\left(\frac{\Delta}{\Gamma}\right) \frac{1}{\Delta^2 +
\Gamma^2 e^{i\lambda}} \, ,
\\                                \label{nrqel}
 \ln \chi_{qel}(\lambda)
&=&
 -\frac{g^2 \T V}{\pi}
 \frac{\Delta^2}{\Delta^2 + \Gamma^2}
 \frac{1}{\Delta^2 + \Gamma^2 e^{i\lambda}}\, ,
 \\                               \label{nriqel}
  \ln \chi_{inel}(\lambda)
&=&
 \frac{g^2 \T (V-\Omega)}{\pi} \, \theta(V-\Omega)
  \\ \nonumber
 &\times&
  \left[
 \frac{\Delta^2}{\Delta^2 + \Gamma^2} -
 \frac{1}{2} \frac{\Delta^2 -
 \Gamma^2 e^{i\lambda}}{\Delta^2 + \Gamma^2 e^{i\lambda}} \right]
 \frac{1}{\Delta^2 + \Gamma^2 e^{i\lambda}} \, .
\end{eqnarray}
In many cases the perturbative FCS of correlated systems is a
linear combination of $e^{i n \lambda}$ terms with some integer
$n$ \cite{muz,cuevas,weisssaleur,AM,reznikov,unitaryFCS}. One
possible explanation is that the corresponding terms describe
transport processes in which the initial dot state is restored
after the tunneling of $n$ electrons. However, $e^{i\lambda}$
enters the above equations \emph{nonlinearly}. This means that
there are processes in which $n$ electrons are necessary in order
to bring the system back to its initial state:  The dot can be
excited by a first transmitted electron, then a number of them can
flow without interaction. The final $n$th electron then deexcites
the system and leaves it in the initial ground state.

An expansion of Eqs.~(\ref{nr1}-\ref{nriqel}) in
$\Delta/\Gamma$, which corresponds to the limit of large
transmission $T_0(\omega) \approx 1$, leads to a series containing
$e^{-i n \lambda}$ ($n > 0$), describing electron
\emph{backscattering} off the dot and thus a reduction of
transmission. In the opposite limit of weak transmission,
$T_0(\omega) \approx 0$, the CGFs can be expanded for small
$\Gamma/\Delta$. One then only encounters
\emph{forward-scattering} terms containing $e^{i n \lambda}$ ($n >
0$) which increase the transmission.


\begin{figure}[t]
  \centering
  \includegraphics[width = 0.48 \textwidth]{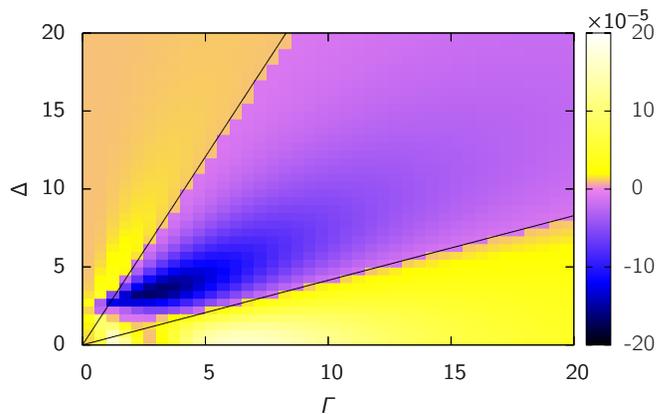}
  \caption{(Color online) Density plot of the inelastic noise contribution $\delta \cumus{Q^2}{inel}/\T$
  for the parameters
  $\Omega = 10 g$ and $V = 10.1g$ as a function of $\Gamma$ and $\Delta$
  measured in units of $g$. Thin lines correspond to the transmissions
  given by Eq.~(\ref{T0}). Dark and bright areas represent negative and positive
  noise correction values, respectively.}
  \label{fig:InelasticTerms}
\end{figure}

Next we would like to discuss the issue of the sign change of the
conductance and noise corrections in vicinity of $V=\Omega$. In
the low-energy sector it is only governed by the inelastic term.
The correction to the current is given by
\begin{align}
\delta I_{inel}
&= -\frac{i}{\T} \frac{d}{d\lambda}\bigg|_{\lambda=0}
 \ln \chi_{inel}(\lambda) \\
&=
 - \frac{g^2 (V-\Omega)}{2 \pi} \, \theta(V - \Omega)
 \frac{\Gamma^2(\Gamma^2 - \Delta^2)}{(\Gamma^2 + \Delta^2)^3} \notag \, .
\end{align}
Here we immediately realize that the sign change occurs at
precisely $\Delta = \pm \Gamma$. As the low-energy transmission
coefficient is given by $T_0(0) = \Gamma^2 /(\Gamma^2 +
\Delta^2)$, the turning point condition is indeed $T_0 = 1/2$.
Interestingly, not only the current, but \emph{all the odd order
cumulants} (due to the inelastic correction) vanish at this point.

The correction to the noise power due to the inelastic tunneling
is given by
\begin{align}
 \delta\cumu{Q^2}{inel}
&=
 -\frac{d^2}{d\lambda^2}\bigg|_{\lambda=0}\ln \chi_{inel}(\lambda) \\
&=
 \frac{g^2 \T (V-\Omega)}{2\pi} \theta(V-\Omega)
 \frac{\Gamma^2(\Gamma^4 - 6 \Gamma^2 \Delta^2 +
 \Delta^4)}{(\Gamma^2 + \Delta^2)^4} \notag \, .
\end{align}
This term changes its sign at the values $\Delta = (\pm 1 \pm
\sqrt{2}) \Gamma$ which correspond to the bare transmission
coefficients given by
\begin{align}            \label{T0}
 T_0 = \frac{1}{2} \left(1 \pm \frac{\sqrt{2}}{2}\right).
\end{align}
In Fig.~\ref{fig:InelasticTerms}, we have plotted the
inelastic noise contribution resulting from the CGF
(\ref{xiinel}) for a bias voltage $V \gtrapprox \Omega$. As soon
as $\Gamma,\Delta > V$ one clearly sees the change of
sign. On the other hand, at low $\Gamma,\Delta < V$, the observed
features are nonuniversal. Since the sign change occurs only for
$V > \Omega$ and the elastic part is featureless around these
critical values of $\Delta$ and $\Gamma$, we believe that the
vanishing of the noise correction should be observable in
experiments.

\begin{figure}[t]
  \centering
  \includegraphics[width = 0.48 \textwidth]{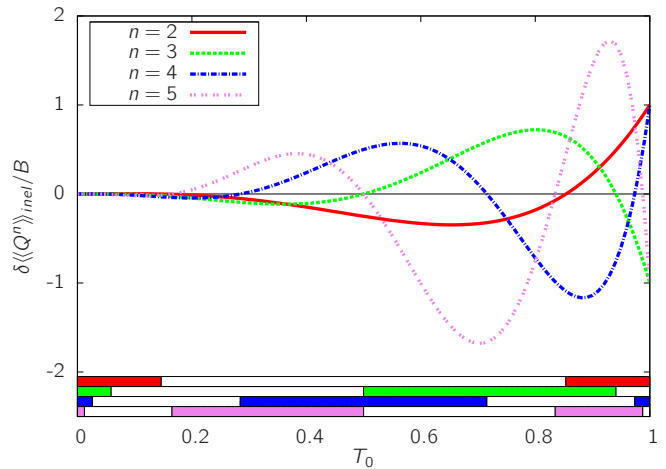}
  \caption{(Color online) Higher-order cumulants $\delta\cumu{Q^n}{inel}$
  normalized to $B = g^2 {\cal T} (V-\Omega)/(2 \pi \Gamma^2)$
   in the low-energy sector as a function of the bare transmission $T_0$.
   The filled rectangles indicate ranges in which the respective cumulant
   is positive. The $n$th order cumulant has $n$ zeros in the interval
   $T_0 \in [0,1]$. The sign changes are
   spaced symmetrically around $T_0=1/2$.}
  \label{fig:UniversalCumulants}
\end{figure}

This behavior proliferates to cumulants of higher orders. In the
$n$th order the turning point condition is given by the zeros of
an $n$th order polynomial. The corresponding solutions for the
transmission coefficient are distributed symmetrically around
$T_0=1/2$. This is demonstrated in
Fig.~\ref{fig:UniversalCumulants}. One also notes that the
cumulants increase drastically towards $T_0 \rightarrow 1$. Hence,
these effects will be observable in junctions of the type used in
Refs.~\cite{djukic05,djukic06,kiguchi08}, where large transmissions $T_0
\approx 1$ have been observed.

To summarize, we have calculated the charge transfer statistics of
a molecular quantum dot with a single fermionic state coupled to a
local Holstein phonon by means of a perturbative expansion in the
electron-phonon coupling. We found that the FCS is the sum of an
adiabatic (mean-field like) term, an elastic part and an inelastic
term. The latter appears as soon as the applied voltage exceeds
the phonon frequency. We found that this inelastic term leaves a
characteristic imprint on the noise power as well as on all higher
order cumulants. We expect these features to become observable in
experiments in the nearest future.

\acknowledgments The authors would like to thank A.~O.~Gogolin and
R.~Egger for many interesting discussions. T.L.S.~is financially
supported by the Swiss NSF and the NCCR Nanoscience. A.K.~is
supported by the DFG grant No.~KO~2235/2 and by the Kompetenznetz
``Funktionelle Nanostrukturen III'' of the Landesstiftung
Baden-W\"urttemberg (Germany).

\bibliography{LocalPolaronPaper}

\end{document}